\documentclass{JFM-FLM_Au}
\usepackage{mathrsfs}
\usepackage{graphicx}% Include figure files
\usepackage{dcolumn}% Align table columns on decimal point
\usepackage{bm}% bold math
\usepackage{subcaption}
\usepackage{xcolor}% For blue
\hypersetup{colorlinks=true, citecolor=blue, linkcolor=blue}

\lefttitle{Nikhil Yewale, Anil Kumar , Vinod Kumar Kadari and Ratul Dasgupta}
\righttitle{Journal of Fluid Mechanics}

\title{A windy sea surface with Stokes waves}

\author{Nikhil Yewale\aff{1}, Anil Kumar\aff{1} , Vinod Kumar Kadari \aff{1} and Ratul Dasgupta\aff{1}}

\affiliation{\aff{1}Department. of Chemical Engineering, Indian Institute of Technology Bombay}

\corresau{Ratul Dasgupta, \email{dasgupta.ratul@gmail.com}}

\begin{document}
\maketitle

\begin{abstract}
Predicting the transition of wind-forced, gravito-capillary surface waves from smooth to corrugated states remains a longstanding problem in nonlinear surface wave mechanics. Solving driven-dissipative, nonlinear potential flow equations we map these waves ($4-13$ cm) onto a Reynolds number -- wave energy phase-space. We identify a transition band separating smooth from corrugated wave states - the wave steepness exhibits a non-monotonic dependence on Reynolds number within this. Subharmonic (in)stability analysis reveals that the fastest-growing mode intensifies corrugations on alternate faces of the carrier wave. Our results offer insights into parasitic capillary wave formation on steep carrier waves and are of interest to ocean remote-sensing.
\end{abstract}


%{\bf MSC Codes }  {\it(Optional)} Please enter your MSC Codes here
%\tableofcontents

\section{Progressive wind-forced waves}
%\noindent\textbf{Progressive wind-forced waves:} 
Observations of sea-wave excitation by wind and the resulting wave spectra, have led to deep scientific questions, frequently impacting ocean engineering. One hundred and fifty-five years after Kelvin's \citep{thomson1871xlvi} theoretical estimate of the threshold speed at which wind raises surface waves, the subject remains extraordinarily active \citep{geva2022excitation, scapin2025momentum,hristov1998wave,KumarAnil2026}. Short waves (a few cm to tens of cm long) have been of particular interest in the study of wind-wave generation. Short, wind-forced, spatially-periodic (but not necessarily sinusoidal), progressive waves propagating at constant speed without change of form have been particularly useful idealisations in the subject. For example, such permanent-form waves have helped rationalise aspects of the wind-forced, sea-wave spectra or their laboratory facsimiles \citep{komen1980nonlinear,lake1978new}. By construction, these nonlinear, progressive waveforms maintain their shape under steady wind-forcing -- the energy input from wind exactly compensating viscous dissipation.

In this study, we model the wind-driven sea surface as comprising such nonlinear waves of permanent form and derive first-principles regime map(s). These accept as input the carrier wavelength and amplitude, and predict the shape, phase-speed and wind stress needed to maintain the waveform, while also predicting whether the wave is smooth or corrugated. In the (non-dimensional) space spanned by the wave-based Reynolds number $Re_{\lambda}$ and wave-energy $\mathscr{E}$, our maps distinguish \textit{smooth} nonlinear carrier waves vis-\`a-vis those which feature corrugations on one (leeward) or both faces (windward and leeward). Functions encapsulating information on wind-forcing $P_{\lambda}(Re_{\lambda},\mathscr{E})$, the Bond number $\alpha_{\lambda}(Re_{\lambda},\mathscr{E})$, the non-linear dispersion relation via the Froude number $F_{\lambda}(Re_{\lambda},\mathscr{E})$, and wave-steepness $\epsilon(Re_{\lambda},\mathscr{E})$ are determined as well. This is done for wavelengths of $4-13$ cm and steepnesses (amplitude times wavenumber) of $10^{-5}-0.37$. In addition to its fundamental aspect, our analysis informs oceanic remote-sensing, where predicting corrugations on carrier waves is considered crucial \citep{plant2017joint}, necessitating empirical estimates.

\section{Smooth and corrugated Stokes waves}
%\noindent\textbf{Smooth and corrugated Stokes waves:} 
Since the inception of the subject by George Stokes, inferring the shape and propagation speed of permanent-form surface-waves from the nonlinear potential flow equations with gravity \citep{stokes1847theory} or surface-tension \citep{wilton1915lxxii}, deducing properties of their limiting forms \citep{crew2016new}, and predicting their stability \citep{benjamin1967disintegration} have persisted as topics of great interest in sciences \citep{smirnova2024water, wilkening2011breakdown,ablowitz2000modulated} and engineering. Sophisticated techniques borrowed from analytical mechanics have generated deep insights \citep{zakharov2009modulation,zakharov1968stability} into the instabilities of such waves, often finding applications in understanding mechanisms of rogue wave formation \citep{toffoli2024observations,onorato2001freak,shukla2006instability}. At the applied end of climate science, knowledge gleaned from numerical studies \citep{iafrati2013modulational,deike2015capillary,kadari2024wave} which employ such waves to initialise Direct Numerical Simulations (DNS), has informed predictions concerning wave-breaking and air entrainment into the upper-sea layer. 
%    \begin{figure}
	%		\includegraphics[scale=0.15]{Images/Fig0.png}
	%		\caption{A wind-driven water pool at Bordeaux, France (Miroir d'eau) featuring short waves. Photo taken in June 2024.}
	%		\label{fig0} 
	%	\end{figure} 
\begin{figure}
	\centering
	\includegraphics[scale=0.25]{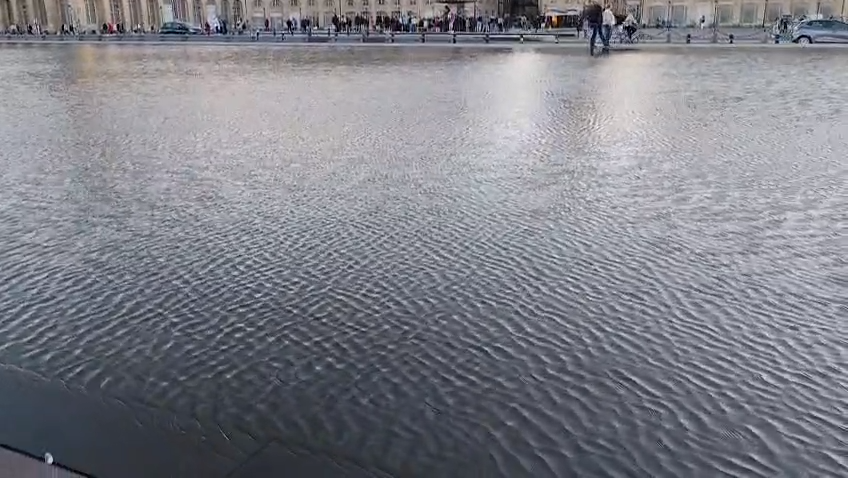}
	\caption{A water pool at Bordeaux, France (Miroir d'eau), a few centimetres deep and featuring wind-generated, transient, short waves. The scale may be estimated from the two standing humans seen in the upper part of the image. Photo taken in June 2024 by RD.}
	\label{fig0}
\end{figure}

At low-enough wind conditions when the air-water surface, while free from wave-breaking may nevertheless still feature waviness (see fig. \ref{fig0} for a commonplace example), corrugations (roughness) are often seen on the surfaces of wind-generated progressive waves e.g. see results in \cite{zhang1995capillary,ebuchi1987fine,hidy1966wind} in the laboratory as well as in the field \citep{cox1957measurements}. Quantifying these corrugations and relating them to the properties of the carrier waves has been of considerable interest: not only towards estimating the momentum transfer across the windy sea-surface via the drag coefficient, but also because this knowledge is necessary for interpreting images of the remotely-sensed ocean. These roughnesses on the faces of carrier waves, at millimetric or even centimetric scales \citep{zhang1995capillary}, can either be due to short, \textit{bound-wave components} \citep{komen1980nonlinear} or \textit{free ripples} \citep{longuet1987propagation,curcic2025revisiting}. The bound-ripples violate their dispersion relation, travelling instead at the phase-speed of their parent carrier wave -- the entire wave profile thus appears steady in the co-moving frame, the Stokes wave being a classic example where bound components appear. In contrast, \textit{free-ripples} travel at their respective phase-speeds and the entire waveform necessarily distorts as it propagates. Recent research has emphasised the role of bound-wave components \citep{plant2017joint} in remote-sensing of ocean waves. These models require estimating the probability of finding bound-waves on longer carrier waves (see eqn.\ B13 in \cite{plant2017joint}). Lacking first-principle theoretical results, fixed empirical estimates of the steepness cutoff beyond which bound components may appear, are often employed. On the other hand, using DNS initialised with a third-order representation of the Stokes wave \textit{without} wind-forcing, \cite{deike2015capillary} reported a sharp boundary in the space of Bond number (gravity to surface tension force) versus carrier-wave steepness. Their regime map delineates the region where parasitic capillary ripples (PCW) may be observed in DNS, from where the wave surface remains smooth with time, albeit it can break in varying modes \citep{deike2015capillary}. 

Our analytical and computational results here, investigate wind-forced, gravito-capillary (GC) waves with viscous dissipation (modified Stokes waves) using the mathematical framework of \citep{dyachenko1996nonlinear,choi1999exact,shelton2025time}. These calculations have been constrained by air-water properties, thereby reducing the dimensionality of the non-dimensional parameter space. Our results severely revise the influential early estimates of \citep{deike2015capillary,plant2017joint} for identifying smooth versus corrugated waves -- wind stress, a crucial ingredient absent in \cite{deike2015capillary}, appears in our formulation. Due to our emphasis on permanent waveforms under wind-forcing, the corrugations in our profiles originate \textit{purely} from bound-wave contributions. We demonstrate that in the same non-dimensional space as \cite{deike2015capillary}, the corrugated regime -- treated thus far as a single entity (with PCWs only on the front face of the carrier wave) -- actually features sub-regimes where carrier waves possess ripples on their leeward face only, and a distinct second sub-regime where these ripples appear on both faces; a transition region links the two sub-regimes. Furthermore, we show that the boundary between the smooth and the corrugated regime, often thought to be a sharp one \citep{deike2015capillary,plant2017joint}, \textit{is not so}, featuring instead an effective finite ``thickness'' due to the wave-steepness not varying monotonically within the transition region. While one and two-sided corrugations on driven Stokes waves have been reported earlier in \cite{shelton2025time} (their fig.\ $2$), the ratio $\dfrac{F_{\lambda}}{Re_{\lambda}\alpha_{\lambda}^{3/4}}=M_{o}^{1/4}$ (Morton number $M_o$) for their fig.\ $2$ varies from $2.06\times10^{-3}-1.65\times10^{-2}$. This ratio depends only on gas-liquid properties, being  $2.29\times10^{-3}$ for air-water. One of the key contributions of this study is the reduction in the dimensionality of the underlying non-dimensional space by constraining calculations to a wind-forced, \textit{air-water} interface. The Morton number is held fixed at the air-water value in all our calculations, unlike \cite{shelton2025time}, and leading to the regime map in fig.\ \ref{fig3} - a key result from our study. We commence analysis with linearised waves in \S\ref{sec:3} followed by \S\ref{sec:4} where nonlinear waves are treated.

\section{Linearised waves ($\epsilon \rightarrow 0$) of permanent form with wind and viscous dissipation}\label{sec:3}

\begin{figure*}[htbp]
\centering
\begin{minipage}[t]{0.48\textwidth}
\centering
\includegraphics[width=\linewidth]{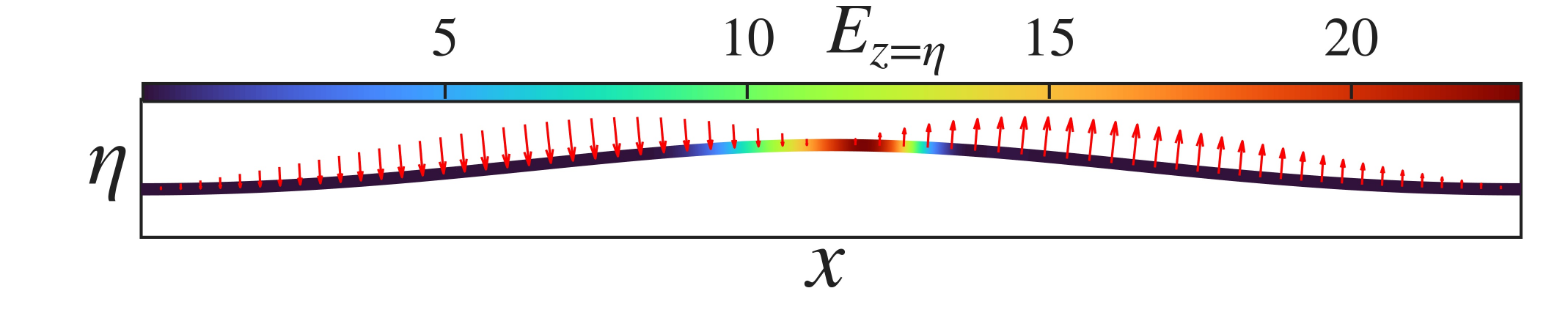}
\captionsetup{width=\linewidth}
\caption{Linearised, \textit{wind-forced} (blowing rightwards) wave of permanent sinusoidal form in a frame co-moving with unit phase speed (non-dimensional); fluid velocity $u(x,z\to-\infty)=-1$. Arrows indicate wind perturbation-pressure \cite{miles1957generation} of form $P_{\lambda}\eta_{x}$ (magnified several orders). Colour contours at $z=\eta(x)$ show viscous dissipation $E = Re_{\lambda}^{-1}(\phi_{xx}^2 + \phi_{zz}^2 + 2\phi_{xz}^2)_{z=\eta}$, $Re_{\lambda}=5.5\times10^4$. Here $\phi(x,z)$ is the steady perturbation velocity potential.}
\label{fig1}
\end{minipage}\hfill
\begin{minipage}[t]{0.48\textwidth}
\centering
\includegraphics[width=\linewidth,trim={0 0.3cm 0 0}, clip]{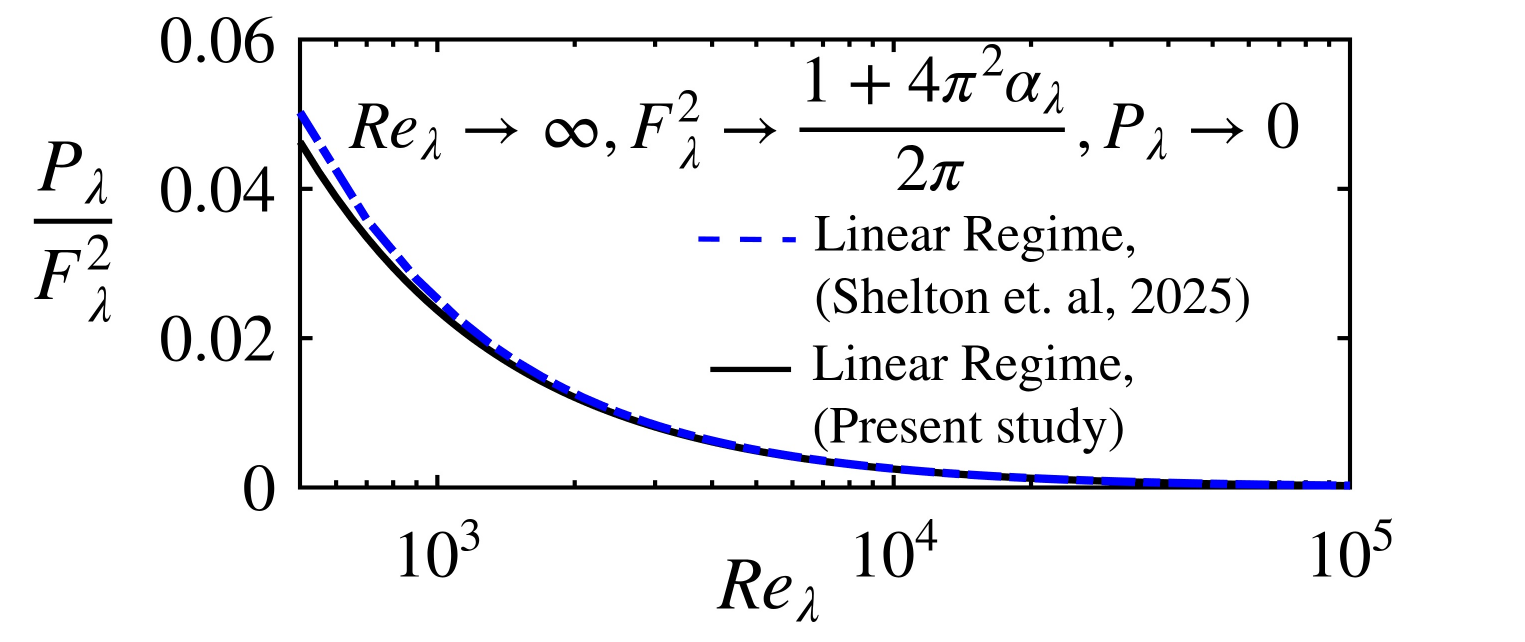}
\captionsetup{width=\linewidth}
\caption{The variation of $P_{\lambda}/F^2_{\lambda}$ with $Re_{\lambda}$ in eqn.\ \ref{eq2}(b) for linearised wind-forced waves of permanent form. The dashed curve is from eqn.\ $2.12$(b) in \cite{shelton2025time}, obtained by neglecting the surface vorticity contribution $\hat{A}_{\hat{x}\hat{z}}$ in eqn.\ \ref{eq1}(c). For $Re_{\lambda} > 10^3$, this term makes a negligible contribution. Expectedly, wind-forcing strength $P_{\lambda}\rightarrow 0$ as $Re_{\lambda}\rightarrow\infty$, see text.}
\label{fig2}
\end{minipage}
\end{figure*}

Commencing with linearised analysis first, consider a (steady and smooth) sinusoidal waveform under wind-forcing, in a frame co-moving with the wave, the latter propagating at its phase-speed $c >0$ with \textit{fixed shape}. Fig.\ \ref{fig1} shows this (sinusoidal) wave in the co-moving frame. The horizontal fluid velocity is $u(x,z\rightarrow-\infty)=-1$, non-dimensionalised by $c$. Given the wavelength $\lambda$, we seek to determine the wind-forcing strength $\hat{p}_w$ and speed $c$ with which the wave can propagate \textit{without} change in form, the wind energy input exactly balancing viscous dissipation.

For this, recall that a flat air-water interface with hydrostatically varying pressure in the water layer (and zero air-pressure) constitutes a trivial background state, about which the Navier-Stokes (NS) equations and boundary conditions may be linearised (see section S2 in Supplementary Material, S.M.). The wave, along with it's accompanying air-pressure forcing as depicted in fig.\ \ref{fig1}, arises as a perturbation on this background state. We use the Helmholtz decomposition for the perturbation velocity i.e.\ $\mathbf{\hat{u}}  = \hat{\bm{\nabla}}\hat{\phi} + \hat{\nabla}\times \hat{\mathbf{A}}$, with $\hat{\phi}$ and $\hat{\mathbf{A}}$ for the irrotational and rotational parts of the perturbed velocity field respectively. The disturbed air-water interface is represented by $\hat{z} = \hat{\eta}(\hat{x})$ (see fig.\ \ref{fig1}, hatted variables are dimensional), with $\hat{\mathbf{A}} = \left[0,0,\hat{A}(\hat{x},\hat{z})\right]$, the wind is assumed to exert \textit{steady} perturbation pressure of the well-known Miles \citep{miles1957generation} form in this frame i.e. $\hat{p}_a(x)=\left(\hat{p}_w / \rho\right)\hat{\eta}_{\hat{x}}$ ($\rho$ density of water, subscript  $\hat{x}$ for partial derivative). Substituting these into the linearised, \textit{steady} NS equations, the equations governing $\hat{\phi}$ and $\hat{A}$ are $\hat{\nabla}^2\hat{\phi} = 0, \quad \hat{\nabla}^2\hat{A} = -\left(\dfrac{c}{\nu}\right)\hat{A}_{\hat{x}}$ with perturbation pressure on the water side $\dfrac{\hat{p}}{\rho} = c\hat{\phi}_{\hat{x}}$ and boundary-conditions at the interface (subscript $\hat{z}$ indicate partial derivative):
%\begin{widetext}
\begin{subequations}\label{eq1}
	\begin{align}
		&-c\hat{\eta}_{\hat{x}} = \hat{\phi}_{\hat{z}} - \hat{A}_{\hat{x}}, \quad 2\hat{\phi}_{\hat{x}\hat{z}} + \hat{A}_{\hat{z}\hat{z}} - \hat{A}_{\hat{x}\hat{x}}=0, \tag{\theequation a,b} \\
		&- c\hat{\phi}_{\hat{x}} - 2\nu\!\left(\hat{\phi}_{\hat{x}\hat{x}} + \hat{A}_{\hat{x}\hat{z}} \right) + g\hat{\eta} - \dfrac{T}{\rho}\hat{\eta}_{\hat{x}\hat{x}} = -\dfrac{\hat{p}_w}{\rho}\hat{\eta}_{\hat{x}}, \tag{\theequation c}
	\end{align}
\end{subequations}
where $\rho,T,g,\nu$ represent water density, air-water surface-tension, acceleration due to gravity and water kinematic viscosity respectively.

Eqns.\ \ref{eq1}(a),(b),(c) evaluated at $\hat{z}=0$ represent the kinematic boundary condition, the vanishing of tangential shear stress and the normal stress conditions, respectively, in the linearised approximation. Employing a Fourier transform along $x$, these equations and boundary conditions are solved in Fourier space. After algebra (See sections S2, S2.1 in S.M.) and non-dimensionalisation, we obtain the following two relations presented in eqns.\ \ref{eq2}(a),(b). These denote the dispersion relation $F_{\lambda} \equiv \dfrac{c}{\sqrt{g\lambda}}$ (also see articles \S$270$,\S$349$ in \cite{lamb1932hydrodynamics}) and the non-dimensional wind-stress $P_{\lambda} \equiv \dfrac{\hat{p}_w}{\rho g \lambda}$ as functions of inverse Bond $\left(\alpha_{\lambda} \equiv \dfrac{T}{\rho g \lambda^2}\right)$ and Reynolds $\left(Re_{\lambda} \equiv \dfrac{c\lambda}{\nu}\right)$ numbers i.e.
\begin{subequations}\label{eq2}
\begin{equation}
F_{\lambda}^2 = \dfrac{1 + 4\pi^2 \alpha_{\lambda}}{Q_r(Re_{\lambda})}, \qquad \dfrac{P_{\lambda}}{F_{\lambda}^2} = \dfrac{1}{2\pi}Q_i(Re_{\lambda}), \tag{\theequation a,b}
\end{equation}
\end{subequations}
where $Q_r$ and $Q_i$ are real and imaginary parts of \mbox{$Q(Re)$}, with
\begin{align}
Q_r(Re_\lambda) &\equiv 2\pi - 16\pi^2 Re_\lambda^{-2}(2\pi - \mathcal{S}_{+}), \label{eq:Qr}\\
Q_i(Re_\lambda) &\equiv 16\pi^2 Re_\lambda^{-1} - 16\pi^2 Re_\lambda^{-2}\,\mathcal{S}_{-}, \label{eq:Qi}\\
\mathcal{S}_{\pm} &\equiv \bigl[\pi\bigl((4\pi^2 + Re_\lambda^2)^{1/2} \pm 2\pi\bigr)\bigr]^{1/2}. \label{eq:S}
\end{align}
Evidently, there are four independent non-dimensional numbers $Re_{\lambda}$, $F_{\lambda}$, $P_{\lambda}$, and $\alpha_{\lambda}$, related by eqns.\ \ref{eq2}(a) and \ref{eq2}(b). Without further constraints these two equations determine, for example, $F_{\lambda}(Re_{\lambda},\alpha_{\lambda}),P_{\lambda}(Re_{\lambda},\alpha_{\lambda})$. However, by constraining the calculation to wind-driven waves (i.e.\ air-water) where $M_{0}^{1/4} \approx 2.29\times 10^{-3}$, we may express $\alpha_{\lambda}$ in terms of $\left(Re_{\lambda},F_{\lambda}\right)$. This shrinks the dimensionality of the space; the (scaled) wind-forcing strength $P_{\lambda}/F_{\lambda}^2$ in eqn.\ \ref{eq2}(b) is a function of $Re_{\lambda}$ only, while eqn.\ \ref{eq2}(a) representing the viscous dispersion relation for wind-forced GC waves, transforms into an implicit relation of the form $f(F_{\lambda},Re_{\lambda})=0$.

Fig.\ \ref{fig2} (solid black curve) presents the variation of (scaled) wind-stress $P_{\lambda}/F_{\lambda}^2$ with $Re_{\lambda}$ (eqn.\ \ref{eq2}(b)). We notice that $Q_{r}(Re_{\lambda}\rightarrow \infty)\rightarrow 2\pi, Q_{i}(Re_{\lambda}\rightarrow \infty)\rightarrow 0$ implying from eqn.\ \ref{eq2}(a), that $F_{\lambda}^2\rightarrow (1+4\pi^2\alpha_{\lambda})/2\pi$; this is simply the inviscid, GC dispersion relation in deep-water without wind. The wind forcing $P_{\lambda}$ vanishes in the same limit i.e., no wind is required to maintain a sinusoidal progressive wave of permanent form, on an inviscid water pool. This far, our waves have been smooth and sinusoidal. Our viscous dispersion relation, eqn. \ref{eq2}(a) will constrain our regime map developed next, for \textit{finite-amplitude} wind-forced waves; these nonlinear waves can feature corrugations unlike the linearised ones seen so far.

\section{{Smooth and corrugated, nonlinear waves}}\label{sec:4}
%\noindent\textbf{Smooth and corrugated, nonlinear waves:}
\begin{figure}
	\centering
	\includegraphics[scale=0.45,trim={0 1cm 0 0}, clip]{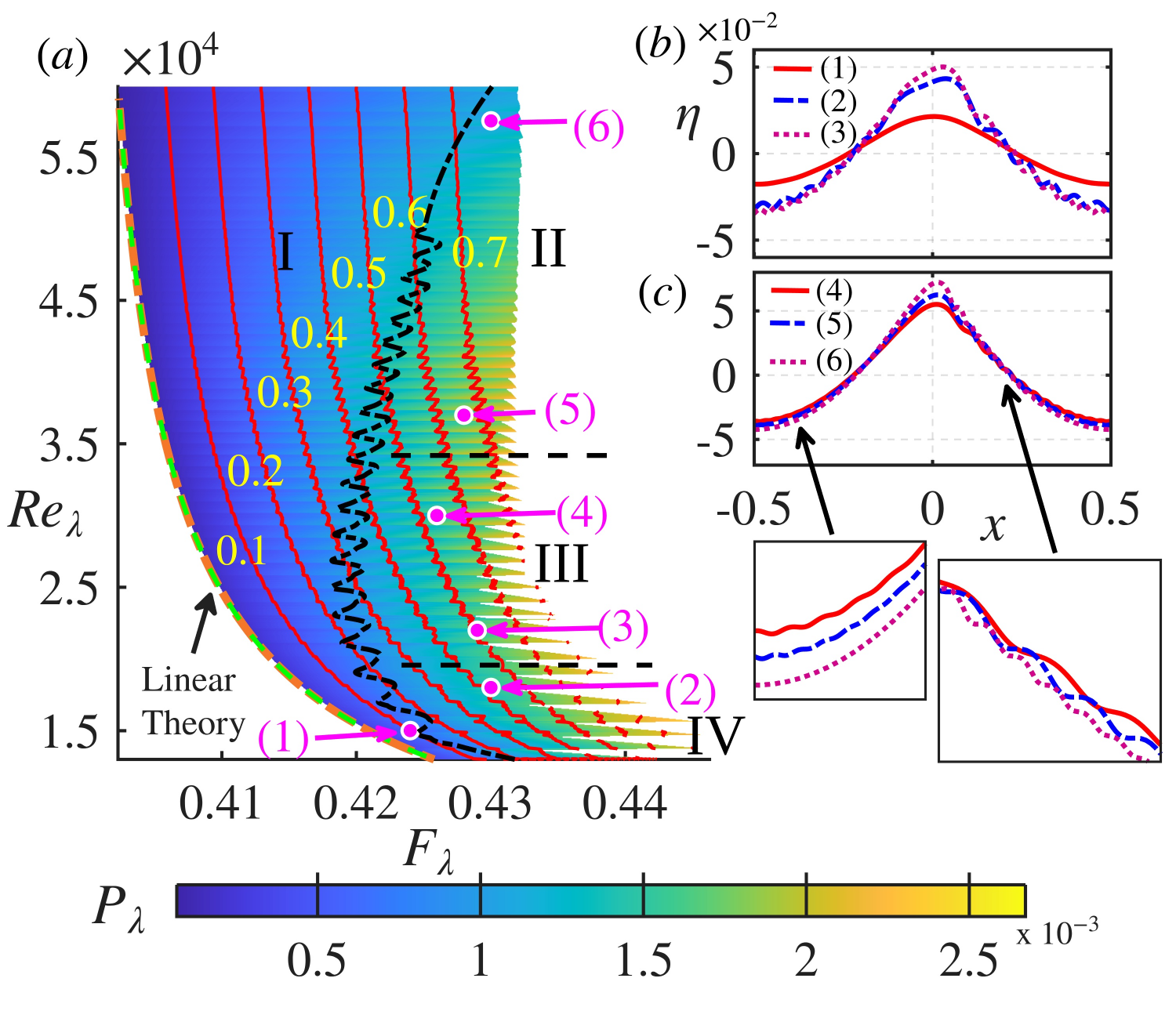}
	\caption{Panel (a) Regime map - Each point on the map corresponds to a GC, wind-forced Stokes wave with dissipation at chosen $(Re_{\lambda},\mathscr{E})$,  red curves ($\mathscr{E}=0.1,0.2\ldots0.7$) are constant energy ($\mathscr{E}$) contours. Background color contours indicate non-dimensional wind-strength $P_{\lambda}(Re_{\lambda},\mathscr{E})$ -- note its small variation across most of the map. The prominent black dashed-dot curve separates smooth (region I, left) from corrugated regimes (II, III, IV, right). Region II: predominantly leeward ripples (right face); IV: windward and leeward ripples; III: transition between II and IV. The prominent black dashed-dot curve is obtained using the procedure described in sec. S6 in the S.M. Panels (b) \& (c) GC wave shapes sampled at the regions indicated as (1)-(6) in panel (a).}
	\label{fig3} 
\end{figure}
\begin{figure}
	\centering
	\includegraphics[scale=0.4,trim={0 0.6cm 0 0}, clip]{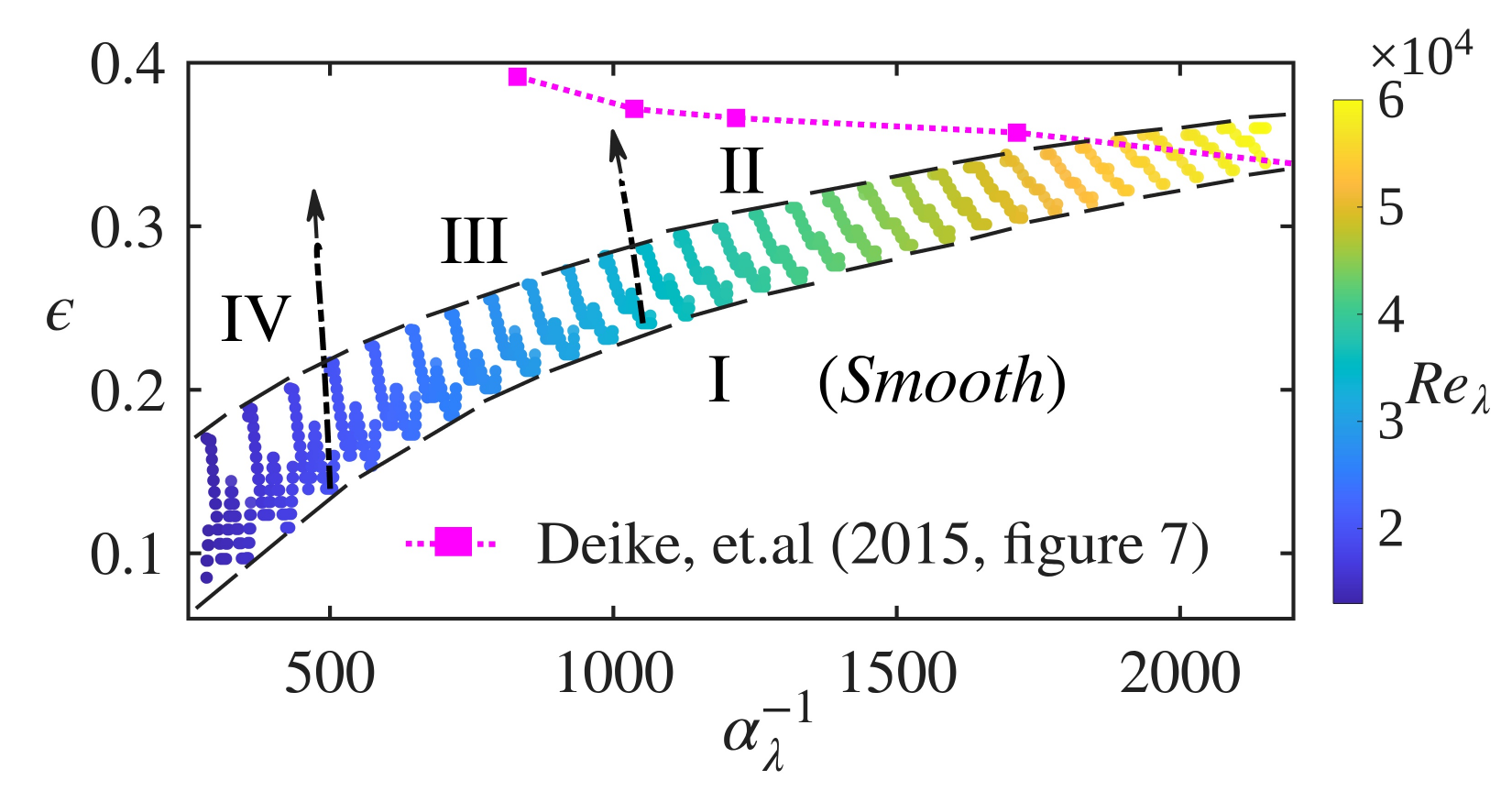}
	\caption{Regime map - The region nomenclatures (I, II, III and IV) are the same as fig.\ \ref{fig3}. The dashed lines straddling the transition region serve as visual guides. The critical Bond $\alpha_{\lambda}^{-1}$ beyond which only smooth waves appear $>2500$ with $Re_{\lambda} \approx 6\times10^4$. Unlike the (fixed) empirically estimated critical steepness in \cite{plant2017joint}, the critical steepness here varies signficantly with $Re_{\lambda}$.}
	\label{fig4}
\end{figure}
Consider now, wind-forced, \textit{nonlinear} waves appearing steady in the co-moving frame. For determining these, we solve the nonlinear equations of potential flow written in the co-moving frame with a vortical contribution to the vertical component of perturbation velocity. This is similar to the linear regime, although the contribution from $\hat{A}_{\hat{x}\hat{z}}$ in eqn.\ \ref{eq1}(c) has been ignored (see caption to fig.\ \ref{fig2} for justification). In addition, we also drop the constraint of zero tangential stress at the interface. A dynamic boundary condition is derived using the nonlinear Bernoulli equation with a (linearised) viscous contribution to normal stress (see eqns S3.1, S3.2 in S.M. based on \cite{dias2008theory}) from the water-side. The kinematic and dynamic boundary conditions are conformally mapped following the approach laid out in \cite{shelton2025time,dyachenko1996nonlinear,choi1999exact}. These equations and accompanying algebra are lengthy; see section S4 in S.M. for details. The resulting nonlinear equations are posed and solved as a root-finding problem in the discretized collocation variable $\xi$ (section S4.2 and S5 in S.M. for details). Note that compared to linearised waves, finite-amplitude waves have an additional descriptor, viz.,\ their amplitude $\hat{a}$, this being indeterminate in linear theory. Consequently, an additional non-dimensional parameter, viz.,\ carrier-wave steepness $\epsilon = \hat{a}\hat{k}$, characterises the finite-amplitude waves in addition to $Re_{\lambda}$. As shown by \cite{shelton2025time}, it is more convenient in numerical calculations to choose this second descriptor as the non-dimensional wave-energy $\mathscr{E}$ instead of its steepness $\epsilon$. By solving the full nonlinear equations with wind-forcing and dissipation (see eqns. S4.6 and S4.7 in S.M.) as per the algorithms elaborated in section S5.3 of the S.M., in the co-moving frame for chosen values of $\left(Re_{\lambda},\mathscr{E}\right)$, we determine the shape of the wind-forced GC wave (and thereby its steepness $\epsilon(Re_{\lambda},\mathscr{E})$) in addition to also finding $P_{\lambda}(Re_{\lambda},\mathscr{E}), F_{\lambda}(Re_{\lambda},\mathscr{E})$ with $\alpha_{\lambda}^{-3/4}=2.29\times10^{-3}Re_{\lambda}F_{\lambda}^{-1}$ (Morton number constraint for air-water). Notably, our procedure differs from \cite{shelton2025time}, where $\alpha_{\lambda}$ was varied independently. 

Fig.\ \ref{fig3}(a) presents our regime map (carrier wavelength range $4-13$ cm), where we distinguish the smooth (region I) from the corrugated regimes (II, III and IV, cf. fig. \ref{fig3}(b),(c)); please see figure caption for details. Eqn.\ \ref{eq2}(a) obtained from linearised analysis earlier, is also plotted in fig.\ \ref{fig3} and indicated as `Linear theory'; this provides a limiting boundary for smooth waves (region I) with steepness $\epsilon\rightarrow 0$. It has been checked that $\epsilon =  \mathbb{O}(10^{-5})$ on this curve. Several benchmarking tests were performed on our nonlinear solutions, recovering, e.g., the Stokes wave without wind (validation section S5.4 in S.M.). We also note from fig. \ref{fig3} that decreasing $Re_{\lambda}$ within the corrugated regime transitions asymmetric, wind-forced GC-waves towards symmetric waveforms. This contrasts the observation by \citet{shelton2025time} in their fig.\ 2 (reproduced in our S.M., fig. $3$), where the opposite transition is noted, apparently due to the absence of the $M_o^{1/4}$ constraint imposed here. 	
Fig.\ \ref{fig4} presents the variation of carrier wave steepness $\epsilon(Re_{\lambda})$ and Bond number $\alpha_{\lambda}^{-1}(Re_{\lambda})$ \textit{on the} black, dashed-dot, wavy transition boundary separating smooth and the rough regimes in fig.\ \ref{fig3}, the value of $Re_{\lambda}$ indicated as color contours. Note that on this wavy transition boundary of fig.\ \ref{fig3}, the Froude number becomes a function of the Reynolds number  $F_{\lambda}(Re_{\lambda})$ i.e. the number of independent variables shrinks to one. Fig.\ \ref{fig4} shows the `band-like' appearance of the zone which separates the smooth from the corrugated regime. 

Magenta symbols in fig.\ \ref{fig4} are extracted from the regime map of \cite{deike2015capillary}, wherein the authors predict one-sided corrugations (PCW) \textit{under} the pink curve. Fig.\ \ref{fig4} significantly revises this estimate: a large smooth region I \textit{also} appears under their pink curve, in addition to region II (correctly predicted by \cite{deike2015capillary}), as well as regions III and IV, which are new results. Fig.\ \ref{fig3} and \ref{fig4} are our regime maps for GC Stokes waves with wind-forcing. To our knowledge, these results in figs. \ref{fig3} and \ref{fig4} constitute first reports of accurate regime maps for finite-amplitude, wind-forced, GC waves of permanent form.

\section{Linear stability}
So far, our inferences have all been from computations of the GC Stokes waves under wind forcing. As the pure gravity Stokes wave (without wind) is known to be susceptible to subharmonic and superharmonic instabilities, it is meaningful to ask if the corrugations seen on wind-forced GC Stokes waves are intensified or ameliorated by a linear instability, when present. Here, we report a sub-harmonic linear stability analysis of wind-forced, finite-amplitide, viscous, GC waves. Towards this, we perturb the base state as:
\begin{equation}
\begin{pmatrix} Z \\ \Phi \end{pmatrix} = \begin{pmatrix} Z_0 \\ \Phi_0 \end{pmatrix} + \tilde{\epsilon}\, e^{\sigma t} \sum_{m=-\infty}^{\infty} \begin{pmatrix} \hat{b}_m \\
\hat{c}_m \end{pmatrix} e^{2\pi i(m+p)\xi}, \quad |\tilde{\epsilon}| \ll 1,
\label{eq:linearstab}
\end{equation}
where $Z_0$ and $\Phi_0$ are the steady-state interface ordinate and velocity potential in the conformally mapped coordinate $\xi$ (see eqn S4.7 in the Supplementary Material (S.M.)  ), $\tilde{\epsilon}\ll 1$ is the perturbation amplitude, $p\in\mathbb{R}$ is the Floquet exponent, $\sigma\in\mathbb{C}$ contains information about growth rates when the base-state is unstable, and $\hat{b}_m$, $\hat{c}_m$ are the Fourier coefficients of the perturbation eigenvector for the interface and velocity potential respectively. \citet{shelton2025time} showed that these waves are superharmonically stable ($p=0$) and we extend their analysis to the sub-harmonic case here. Our analysis shows that similar to the classical Stokes wave (without capillarity, wind-forcing and dissipation cf. \cite{LH1978Sub}), wind-forced viscous gravity-capillary Stokes waves also exhibit \textit{Benjamin--Feir-like} instability.
\begin{figure}[h]
	\centering
	\includegraphics[scale=0.34, trim={0.5cm 0.2cm 0.4cm 0}, clip]{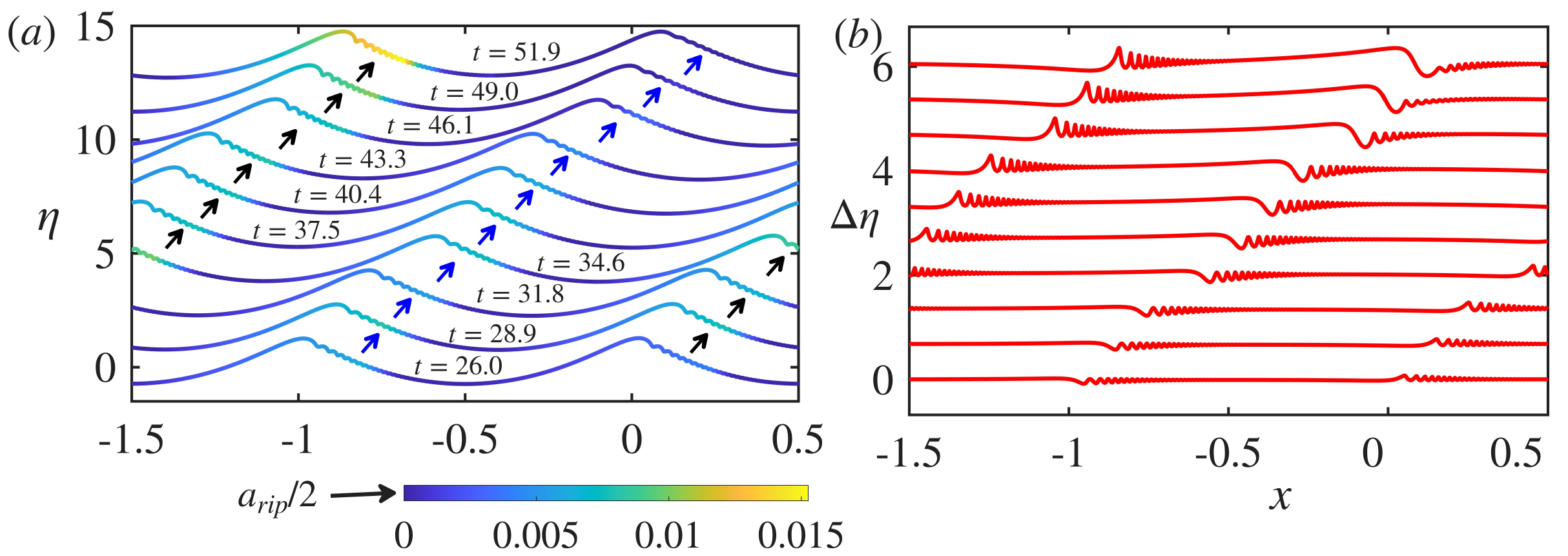}
	\caption{(a) Time evolution of the (non-dimensional) free-surface perturbed by the fastest-growing subharmonic mode. Colour denotes local ripple amplitude (half of peak-to-trough distance $< 0.005\lambda$, $\lambda$ being the carrier wavelength). Blue arrows: ripples weakening with time; black arrows: ripples intensifying. (b) Time evolution of the imposed perturbation eigenmode, showing constructive/destructive interference with the base state that drives the observed modulation of the capillary ripples. The time instances of the perturbation profiles are the same in sequence from bottom to top, as that of panel (a)}
	\label{fig5} 
\end{figure}

Panel~(a) of fig.\ \ref{fig5} shows the nonlinear time evolution of a GC Stokes wave, chosen from region II of our earlier fig.\ \ref{fig3} (with one-sided corrugations). This base-state is representative of steep sea waves with  $\epsilon=0.366$, $F_{\lambda}=0.4316$ and $Re_{\lambda}=5.75\times10^4,$ $\, \alpha_{\lambda}=4.94\times10^{-4}$, $P_{\lambda}=9.28\times10^{-4}$. The base-state turns out to be linearly \textit{unstable} to sub-harmonic perturbations (i.e. perturbations longer than the base state). For generating this figure, the base state has been perturbed using the fastest-growing subharmonic ($p=0.5$) eigenmode ($\sigma = 0.0869 - 1.131\,i$) obtained from linear stability analysis. 

This is then time-evolved using the full nonlinear, time-dependent eqns. (i.e. eqns. in conformally mapped form; see S.M. eqns.~S4.4,~S4.5). Reminiscent of the Benjamin-Feir instability (without wind), the long-time evolution (about fifty-one wave periods in panel (a), fig. \ref{fig5}) of the perturbed waveform reveals alternate crest sharpening and broadening --- a known feature of modulational instability under long-wavelength perturbations. This instability now manifests on the forward-face ripples of the base-state, as alternating weakening and strengthening with time. Panel~(b) of fig.~\ref{fig5} shows the corresponding evolution of the eigemode, confirming that constructive/destructive interference of this with the base state drives the observed modulation of the capillary ripples. Details of the generalised eigenvalue problem formulation and numerical implementation are given in the section S7 of Supplementary Material (S.M.). Further details of the linear instability, its long-term evolution including the stability analysis of background states with two-sided corrugations, will be presented in a forthcoming publication.	

\section{Conclusion}
We have analysed an idealised, steady, wind-driven sea-surface, comprising GC Stokes waves accounting for laminar dissipation. These are computed accurately using the full nonlinear potential flow equations with forcing and dissipation in the linearised as well as nonlinear regimes. Regime maps, presented for the first time here, show that the region separating smooth and corrugated carrier wave-states has for all practical purposes, a `band-like' appearance in the steepness versus Bond number plane (fig. \ref{fig4}), owing to multi-valuedness of carrier steepness $\epsilon(Re_{\lambda})$ within this region. The limiting steepness beyond which corrugations are expected can be read off from fig.\ \ref{fig4}, along with the corresponding wind-stress from fig.\ \ref{fig3}(a). Within the corrugated regime of fig. \ref{fig3}(a), decreasing $Re_{\lambda}$ transitions asymmetric GC waves towards waveforms with both side corrugations in a nearly symmetric manner -- opposite to the trend reported in \cite{shelton2025time}; this is an apparent consequence of fixing Morton number $M_o^{1/4}$ at the air-water value in our study. Linear subharmonic instability results for region II of fig. \ref{fig3}(a) (one-sided ripples) indicate that the corrugations on alternate waves are intensified with time. Specifically, linear stability analysis of finite-amplitude waveforms with wind and dissipation, reveals a Benjamin--Feir-like instability: the fastest-growing subharmonic mode drives alternate crest sharpening and broadening, manifesting as alternating intensification and weakening of the capillary ripples via constructive and destructive interference, on the forward face of the base state. These results have obvious ramifications for remote-sensing as for the physics of nonlinear, wind-driven waves.

% \cite{mundy_metaphysics_1987,Field1980,Field1980y,Field1980z,field_can_1984,PerryForthcoming-PERPEQ,Hoelder1901,eddon_fundamental_????,Arntzenius2012}
%
\begin{bmhead}[Acknowledgements]
	We thank Prof. Rama Govindarajan and Prof. Itamar Procaccia for critical comments on the study.
\end{bmhead}
\begin{bmhead}[Funding]
	We acknowledge support from DST-SERB (Govt. India)
	grants MTR/2019/001240, CRG/2020/003707, SPR/2021/000536, MoE-STARS/STARS2/2023-0595. The doctoral tenures of NY and VK,
	supported by the Prime Ministers Research Fellowship,Ministry of Education, Govt. of India and Ansys Inc., respectively are sincerely acknowledged.
\end{bmhead}
\begin{bmhead}[Declaration of interests]
	The authors report no conflict of interest.
\end{bmhead}
\bibliographystyle{jfm}
\bibliography{jfm}

@article{wilton1915lxxii,
	title={LXXII. On ripples},
	author={Wilton, JR},
	journal={The London, Edinburgh, and Dublin Philosophical Magazine and Journal of Science},
	volume={29},
	number={173},
	pages={688--700},
	year={1915},
	publisher={Taylor \& Francis}
}

@article{LH1978Sub,
	author = {Longuet-Higgins, Michael Selwyn},
	title = {The instabilities of gravity waves of finite amplitude in deep water II. Subharmonics},
	journal = {Proceedings of the Royal Society of London. A. Mathematical and Physical Sciences},
	volume = {360},
	number = {1703},
	pages = {489-505},
	year = {1978},
	month = {04},
	issn = {0080-4630},
	doi = {10.1098/rspa.1978.0081},
	url = {https://doi.org/10.1098/rspa.1978.0081},
	eprint = {https://royalsocietypublishing.org/rspa/article-pdf/360/1703/489/62947/rspa.1978.0081.pdf},
}

@article{stokes1847theory,
	title={On the theory of oscillatory waves},
	author={Stokes, George Gabriel},
	journal={Trans. Cam. Philos. Soc.},
	volume={8},
	pages={441--455},
	year={1847}
}

@article{crew2016new,
	title={{New singularities for Stokes waves}},
	author={Crew, Samuel C and Trinh, Philippe H},
	journal={Journal of Fluid Mechanics},
	volume={798},
	pages={256--283},
	year={2016},
	publisher={Cambridge University Press}
}

@article{benjamin1967disintegration,
	title={The disintegration of wave trains on deep water Part 1. Theory},
	author={Benjamin, T Brooke and Feir, James E},
	journal={Journal of Fluid Mechanics},
	volume={27},
	number={3},
	pages={417--430},
	year={1967},
	publisher={Cambridge University Press}
}

@article{lake1978new,
	title={A new model for nonlinear wind waves. Part 1. Physical model and experimental evidence},
	author={Lake, Bruce M and Yuen, Henry C},
	journal={Journal of Fluid Mechanics},
	volume={88},
	number={1},
	pages={33--62},
	year={1978},
	publisher={Cambridge University Press}
}

@article{deike2015capillary,
	title={Capillary effects on wave breaking},
	author={Deike, Luc and Popinet, Stephane and Melville, W Kendall},
	journal={Journal of Fluid Mechanics},
	volume={769},
	pages={541--569},
	year={2015},
	publisher={Cambridge University Press}
}

@article{iafrati2013modulational,
	title={Modulational Instability, Wave Breaking, and Formation of Large-Scale Dipoles in the Atmosphere},
	author={Iafrati, A and Babanin, Alexander and Onorato, Miguel},
	journal={Physical Review Letters},
	volume={110},
	number={18},
	pages={184504},
	year={2013},
	publisher={APS}
}

@article{KumarAnil2026,
	title={{Waves in a shear flow: transition between the KH, Holmboe and Miles instability}},
	volume={1036},
    journal={Journal of Fluid Mechanics},
	author={Kumar, Anil and Ravichandran, S. and Dasgupta, Ratul}, year={2026}, pages={A10}
}

@article{kadari2024wave,
	title={Wave breaking: spilling and plunging},
	author={Kadari, Vinod Kumar and Dhote, Yashika and Kumar, Anil and Yewale, Nikhil and Dasgupta, Ratul},
	journal={The European Physical Journal Special Topics},
	volume={233},
	number={8},
	pages={1685--1694},
	year={2024},
	publisher={Springer}
}

@book{lamb1932hydrodynamics,
	author    = {Lamb, Horace},
	title     = {Hydrodynamics},
	edition   = {6th},
	publisher = {Cambridge University Press},
	address   = {Cambridge},
	year      = {1932},
	isbn      = {9780521458689}
}

@article{wilkening2011breakdown,
	title={Breakdown of self-similarity at the crests of large-amplitude standing water waves},
	author={Wilkening, Jon},
	journal={Physical Review Letters},
	volume={107},
	number={18},
	pages={184501},
	year={2011},
	publisher={APS}
}

@article{hidy1966wind,
	title={Wind action on water standing in a laboratory channel},
	author={Hidy, George M and Plate, Erich J},
	journal={Journal of Fluid Mechanics},
	volume={26},
	number={4},
	pages={651--687},
	year={1966},
	publisher={Cambridge University Press}
}

@article{dyachenko1996nonlinear,
	title={Nonlinear dynamics of the free surface of an ideal fluid},
	author={Dyachenko, Alexander I and Zakharov, Vladimir E and Kuznetsov, Evgenii A},
	journal={Plasma Physics Reports},
	volume={22},
	number={10},
	pages={829--840},
	year={1996},
	publisher={Pleiades Publishing, Ltd.(Плеадес Паблишинг, Лтд)}
}

@article{dias2008theory,
	title={Theory of weakly damped free-surface flows: a new formulation based on potential flow solutions},
	author={Dias, Frederic and Dyachenko, Alexander I and Zakharov, Vladimir E},
	journal={Physics Letters A},
	volume={372},
	number={8},
	pages={1297--1302},
	year={2008},
	publisher={Elsevier}
}

@article{choi1999exact,
	title={Exact evolution equations for surface waves},
	author={Choi, Wooyoung and Camassa, Roberto},
	journal={Journal of engineering mechanics},
	volume={125},
	number={7},
	pages={756--760},
	year={1999},
	publisher={American Society of Civil Engineers}
}

@article{miles1957generation,
	title={On the generation of surface waves by shear flows},
	author={Miles, John W},
	journal={Journal of Fluid Mechanics},
	volume={3},
	number={2},
	pages={185--204},
	year={1957},
	publisher={Cambridge University Press}
}

@article{longuet1987propagation,
	title={The propagation of short surface waves on longer gravity waves},
	author={Longuet-Higgins, MS},
	journal={Journal of Fluid Mechanics},
	volume={177},
	pages={293--306},
	year={1987},
	publisher={Cambridge University Press}
}

@article{plant2017joint,
	title={A joint active/passive physical model of sea surface microwave signatures},
	author={Plant, William J and Irisov, Vladimir},
	journal={Journal of Geophysical Research: Oceans},
	volume={122},
	number={4},
	pages={3219--3239},
	year={2017},
	publisher={Wiley Online Library}
}

@article{zhang1995capillary,
	title={Capillary--gravity and capillary waves generated in a wind wave tank: observations and theories},
	author={Zhang, Xin},
	journal={Journal of Fluid Mechanics},
	volume={289},
	pages={51--82},
	year={1995},
	publisher={Cambridge University Press}
}

@book{cox1957measurements,
	title={Measurements of slopes of high-frequency wind waves},
	author={Cox, Charles S and others},
	year={1957},
	publisher={Scripps Institution of Oceanography}
}

@article{ebuchi1987fine,
	title={Fine structure of laboratory wind-wave surfaces studied using an optical method},
	author={Ebuchi, Naoto and Kawamura, Hiroshi and Toba, Yoshiaki},
	journal={Boundary-Layer Meteorology},
	volume={39},
	number={1},
	pages={133--151},
	year={1987},
	publisher={Springer}
}

@article{thomson1871xlvi,
	title={{XLVI}. {Hydrokinetic solutions and observations}},
	author={Thomson, William},
	journal={The London, Edinburgh, and Dublin Philosophical Magazine and Journal of Science},
	volume={42},
	number={281},
	pages={362--377},
	year={1871},
	publisher={Taylor \& Francis}
}

@article{scapin2025momentum,
	title={Momentum fluxes in wind-forced breaking waves},
	author={Scapin, Nicol{\`o} and Wu, Jiarong and Farrar, J Thomas and Chapron, Bertrand and Popinet, St{\'e}phane and Deike, Luc},
	journal={Journal of Fluid Mechanics},
	volume={1009},
	pages={A20},
	year={2025},
	publisher={Cambridge University Press}
}

@article{geva2022excitation,
	title={Excitation of initial waves by wind: A theoretical model and its experimental verification},
	author={Geva, Meital and Shemer, Lev},
	journal={Physical Review Letters},
	volume={128},
	number={12},
	pages={124501},
	year={2022},
	publisher={APS}
}

@article{hristov1998wave,
	title={Wave-coherent fields in air flow over ocean waves: Identification of cooperative behavior buried in turbulence},
	author={Hristov, Tihomir and Friehe, Carl and Miller, Scott},
	journal={Physical Review Letters},
	volume={81},
	number={23},
	pages={5245},
	year={1998},
	publisher={APS}
}

@article{toffoli2024observations,
	title={Observations of rogue seas in the southern ocean},
	author={Toffoli, Alessandro and Alberello, Alberto and Clarke, Hans and Nelli, Filippo and Benetazzo, Alvise and Bergamasco, Filippo and Ntamba, B Ntamba and Vichi, Marcello and Onorato, Miguel},
	journal={Physical Review Letters},
	volume={132},
	number={15},
	pages={154101},
	year={2024},
	publisher={APS}
}

@article{ablowitz2000modulated,
	title={Modulated periodic Stokes waves in deep water},
	author={Ablowitz, MJ and Hammack, J and Henderson, D and Schober, CM},
	journal={Physical Review Letters},
	volume={84},
	number={5},
	pages={887},
	year={2000},
	publisher={APS}
}

@article{smirnova2024water,
	title={Water-wave vortices and skyrmions},
	author={Smirnova, Daria A and Nori, Franco and Bliokh, Konstantin Y},
	journal={Physical Review Letters},
	volume={132},
	number={5},
	pages={054003},
	year={2024},
	publisher={APS}
}

@article{shukla2006instability,
	title={Instability and evolution of nonlinearly interacting water waves},
	author={Shukla, Padma K and Kourakis, Ioannis and Eliasson, Bengt and Marklund, Mattias and Stenflo, Lennart},
	journal={Physical Review Letters},
	volume={97},
	number={9},
	pages={094501},
	year={2006},
	publisher={APS}
}

@article{onorato2001freak,
	title={Freak waves in random oceanic sea states},
	author={Onorato, Miguel and Osborne, Alfred R and Serio, Marina and Bertone, Serena},
	journal={Physical Review Letters},
	volume={86},
	number={25},
	pages={5831},
	year={2001},
	publisher={APS}
}

@article{curcic2025revisiting,
	title={Revisiting the hydrodynamic modulation of short surface waves by longer waves},
	author={Curcic, Milan},
	journal={Journal of Fluid Mechanics},
	volume={1015},
	pages={A20},
	year={2025},
	publisher={Cambridge University Press}
}

@article{shelton2025time,
	title={Time-dependent nonlinear gravity--capillary surface waves with viscous dissipation and wind forcing},
	author={Shelton, Josh and Milewski, Paul and Trinh, Philippe H},
	journal={Journal of Fluid Mechanics},
	volume={1003},
	pages={A13},
	year={2025},
	publisher={Cambridge University Press}
}

@article{zakharov1968stability,
	title={Stability of periodic waves of finite amplitude on the surface of a deep fluid},
	author={Zakharov, Vladimir E},
	journal={Journal of Applied Mechanics and Technical Physics},
	volume={9},
	number={2},
	pages={190--194},
	year={1968},
	publisher={Springer}
}

@article{komen1980nonlinear,
	title={Nonlinear contributions to the frequency spectrum of wind-generated water waves},
	author={Komen, GJ},
	journal={Journal of Physical Oceanography},
	volume={10},
	number={5},
	pages={779--790},
	year={1980}
}

@article{zakharov2009modulation,
	title={Modulation instability: the beginning},
	author={Zakharov, Vladimir E and Ostrovsky, Lev A},
	journal={Physica D: Nonlinear Phenomena},
	volume={238},
	number={5},
	pages={540--548},
	year={2009},
	publisher={Elsevier}
}

%%%%%%%%%%%%%%%%%%%%%%%%%%%%%%%%%%%%%%%%%%%%%%%%%%%%%%%%%%%%%%%%%%%%%%%%%%%%%%%%%%%%%%%%%%%%%%%%%%%%%%%%%%%%%%%%%%%%%%%
%%%%%%%%%%%%%%%%%%%%%%%%%%%%%%%%%%%%%%%%%%%%%%%%%%%%%%%%%%%%%%%%%%%%%%%%%%%%%%%%%%%%%%%%%%%%%%%%%%%%%%%%%%%%%%%%%%%%%%%

% \backsection[Supplementary data]{\label{SupMat}Supplementary material and movies are available at \\https://doi.org/10.1017/jfm.2019...}
%

% \backsection[Data availability statement]{The data that support the findings of this study are openly available in [repository name] at http://doi.org/[doi], reference number [reference number]. See JFM's \href{https://www.cambridge.org/core/journals/journal-of-fluid-mechanics/information/journal-policies/research-transparency}{research transparency policy} for more information}
%
% \backsection[Author ORCIDs]{Authors may include the ORCID identifers as follows.  F. Smith, https://orcid.org/0000-0001-2345-6789; B. Jones, https://orcid.org/0000-0009-8765-4321}
%
% \backsection[Author contributions]{Authors may include details of the contributions made by each author to the manuscript'}

%\appendix
%\begin{appen}
%\renewcommand{\theHsection}{appendix.\arabic{section}}
%
%\section{Technical details}\label{appA}
%In order not to disrupt the narrative flow, purely technical material may be included in the appendices. This material should corroborate or add to the main result and be essential for the understanding of the paper. It should be a small proportion of the paper and must not be longer than the paper itself.
%
%\end{appen}\clearpage
%

\end{document}